\documentclass[fleqn,10pt]{wlscirep}
\usepackage[utf8]{inputenc}
\usepackage[T1]{fontenc}
\usepackage{amsmath}
\usepackage{amsfonts}
\usepackage{amssymb}
\usepackage{bm}

\title{Spinor superfluid currents of exciton-polaritons on a split-ring}

\author[1,2,*]{Sevak Demirchyan}
\author[3,2*]{Igor Chestnov}
\author[1,4]{Kirill Kondratenko}
\author[1,5,6,7*]{Alexey Kavokin}
\affil[1]{Russian Quantum Center, Skolkovo IC, Bolshoy Bulvar 30, bld. 1, Moscow 121205, Russia}
\affil[2]{Department of Physics and Applied Mathematics, Vladimir State University, Gorkii Street 87, Vladimir 600000, Russia}
\affil[3]{Department of Physics and Engineering, ITMO University, St. Petersburg, 197101, Russia}
\affil[4]{Moscow Institute of Physics and Technology,
9 Institutskiy per., Dolgoprudny, Moscow Region, 141701, Russian Federation}
\affil[5]{Key Laboratory for Quantum Materials of Zhejiang Province, School of Science, Westlake University, 18 Shilongshan Road, Hangzhou 310024, Zhejiang Province, China}
\affil[6]{Institute of Natural Sciences, Westlake Institute for Advanced Study, 18 Shilongshan Road, Hangzhou 310024, Zhejiang Province, China}
\affil[7]{Spin Optics Laboratory, St. Petersburg State University, Ul’anovskaya 1, Peterhof, St. Petersburg, 198504, Russia}

\affil[*]{sevakdemirchyan@gmail.com; igor\_chestnov@mail.ru; a.kavokin@westlake.edu.cn}

\begin{abstract}
Recently, split-ring bosonic condensates of exciton polaritons have been proposed for realisation of qubits. We formulate an analytical model of a polariton condensate in a one-dimensional ring split by a delta-function potential. The persistent current is stopped by the potential defect embedded in a scalar condensate of non-interacting particles, however, it reappears in the presence of an up-critical non-linearity. The nonlinear supercurrents are characterised by fractional orbital momenta while their eigenfunctions may feature grey or dark solitons. The coupling between the spin and the orbital angular momentum of a spinor condensate affects persistent currents crucially. We show that the spin-orbit interaction (SOI) induces the net superfluid current in the condensate with a spatially uniform spin polarisation. We also find solutions where the SOI-induced currents of two spin components of the condensate propagate in opposite directions. The corresponding states of spinor condensates obey a combined  mirror reflection and the spin-flip symmetry. 
\end{abstract}
\begin{document}

\flushbottom
\maketitle

\thispagestyle{empty}

\section*{Introduction}

Because of the recent progress in nanotechnology, polaritonics became an attractive playground for studying the fundamental aspects of quantum \cite{Sanvitto}, coherent \cite{Scnheider} and nonlinear \cite{Carusotto} phenomena occurring due to the interaction of light with matter, and finds its application in various fields of quantum and atomic optics, as well as for quantum information processing \cite{KavokinNature,Xue}. Historically exciton polaritons (hereafter polaritons) were first predicted and described by Hopfield \cite{Hopfield} and Agranovich \cite{Agranovich}, though their findings were preceded by the work of Pekar \cite{Pekar}, who was the first to describe the change in the exciton energy spectrum due to coupling with light in terms of additional waves. The polaritons in semiconductor microcavities were first experimentally evidenced by Weisbuch et al.\cite{Weisbuch}. 

Polaritons attract a lot of attention because of their fascinating coherent properties, namely, due to the ability of sustaining macroscopic coherence in the condensed phase \cite{Richard,Kasprzak,Christopoulos}. Because of the strong polariton-polariton interaction stemming from the scattering of excitons, polariton condensates demonstrate a superfluid behaviour which is manifested by the formation of persistent currents \cite{Carusotto}.  The persistent flow of polaritons is clearly evidenced in the annular geometry \cite{Lukoshkin2017} which can be realised with the use of various approaches including sculpting the pump beam with a spatio-optical modulator \cite{Schnars2005}, etching of ring-channels \cite{Mukherjee2019} or use of pillar microcavities \cite{Kalevich2014,Kalevich2015}. Recently, some of us proposed to exploit the quantum degree of freedom of the polariton current trapped in an annular geometry to encode the state of the quantum bit of information \cite{Xue}. The similar proposals have been made in a field of atomtronics \cite{Amico2021} and with the atomic BEC \cite{Solenov2011}. The practical implementation of the macroscopic coherence associated with the persistent currents requires the tools for a fine tuning of their properties.

An efficient approach to this problem implies the use of the static or moving potential embedded in an ideal ring. The impact of a point-like defect (a weak link) on the circular superfluid currents was investigated in ring-shaped atomic BECs containing a weak link \cite{EckelNature2014,EckelPRX2014} and in the superfluid helium flowing through the pinhole \cite{Sato2011}. Although a superfluid can pass through the defect provided that its speed is less than the speed of sound, the defect affects the net circulation and the energy of the superfluid state. In a stark contrast to these systems, polariton condensates are characterized by spinor rather than the scalar order parameters. Polaritons inherit the pseudospin degree of freedom from two possible orientations of the spin of optically active excitons in zinc-blend semiconductor quantum wells (QWs) such as $\mathrm{GaAs/Al_x Ga_{1-x}As}$ QWs  \cite{KKavokin}. This peculiarity of polariton condensates is responsible for many unique phenomena such as the spin Meissner effect \cite{Rubo} etc.

The use of the two-component spinor order parameter allows to exploit the coupling between the spin and orbital degrees of freedom. In a polariton condensate, the spin-momentum interaction naturally appears due to the momentum-dependent energy splitting of the modes of different polarisations. This phenomenon stems from the longitudinal-transversal splitting of excitonic states \cite{Maialle,Tassone} and the splitting between transverse electric (TE) and transverse magnetic (TM) photon modes of the microcavity. In this case, spin-orbit coupling is quadratic in polariton momentum \cite{Panzarini} while the effective magnetic field acting upon the polariton pseudospin (Stokes vector) is described by second angular harmonics. An impact of this type of SOI on the spin textures of the ring-shaped polariton condensate was  studied in Refs.\cite{Gulevich2016,Zezyulin,Mukherjee2021}.

In this paper, however, we consider another instructive example of spin-orbit coupling. Seeking for the tool of manipulation of the orbital currents of polaritons, we consider a simplest case of SOI which locks polariton angular momentum to its pseudospin in a way that splits linearly polarized polaritons in energy. This type of coupling resembles Rashba spin-orbit interaction experienced by a charged carrier in a quantum ring in the presence of radial electric field \cite{Fomin}. Besides, we investigate the impact of a weak link on the persistent current of a spinor superfluid. In particular, we focus on the possibility of using a combination of the SOI and the defect for a fine tuning of the spinor supercurrent properties. Our analysis reveals various families of the persistent current states which arise due to the interplay between the SOI and polariton-polariton interactions.

\section*{Results}
\subsection*{Eigenstates of an ideal spinor condensate in the split ring geometry}
\begin{figure}
\includegraphics[width=\linewidth]{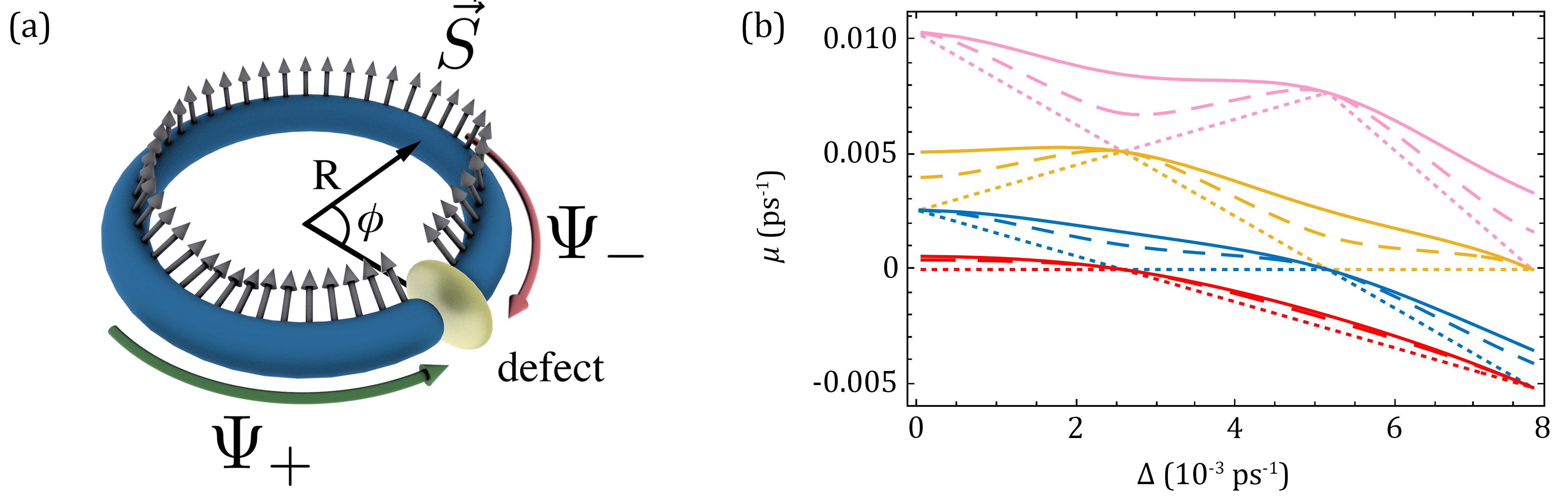}
\caption{(\textbf{a}) Schematic showing split-ring spinor condensate. The green and the red arrows indicate the presence of the circular currents of the spinor superfluid, while grey arrows show the azimuthal distribution of the polariton  spin (pseudospin) vector $\vec{S}$. (\textbf{b}) The $\Delta$-dependence of the energy spectrum for different heights of the delta-potential. The four lowest energy levels are shown with different colors. Dotted lines correspond to the empty ring $\beta=0$, dashed lines to $\beta=2.5\hbar/2m^\star$ while solid lines to $\beta=10\hbar/2m^\star$. The real values of the parameters are $m_{\rm pol}= 10^{-4}m_{\rm fe}$, $R=15$ $\mu$m, where $m_{\rm fe}$ is the free electron mass.}
\label{fig:1}
\end{figure}

We consider a system sketched in Fig.~\ref{fig:1}a. A two-component condensate is localized on a thin ring of the radius $R$. It is characterised by the spinor wave function $\bm{\Psi}=(\Psi_+,\Psi_-)^\intercal$ governed by the following Schr\"{o}dinger equation:
\begin{equation} \label{eq:1}
i\hbar \partial_t\bm{\Psi}= \left[\hat{\mathcal{H}}_{\rm 0} + \hat{\mathcal{H}}_{\rm SOI} + \hat{\mathcal{H}}_{\rm nl} \right] \bm{\Psi} =\left[ {\hat{{h}}_0}\hat{\sigma}_0 - \Delta\hat{\sigma}_y \hat{L}_z + \hat{\mathcal{H}}_{\rm nl} \right]\bm{\Psi},
\end{equation}
where $\hat{{{h}}}_0=-\frac{\hbar^2}{2m^\star}\partial_{\phi\phi}^2 + \hbar \beta \delta(\phi)$, $\hat{\sigma}_0$ is the unity matrix and $\hat{\sigma}_{x,y,z}$ are the Pauli matrices. 
Here $\phi$ is the polar angle, $m^\star$ is the effective polariton mass characterising its azimuthal motion. In the case of a thin ring with the width $w \ll R$ one can connect $m^\star$ with the 2D polariton mass by $m^\star = m_{\rm pol}R^2$. $\beta$ is the strength of the potential defect. The linear coupling between spin-components is described by the term $\hat{\mathcal{H}}_{\rm SOI} = - \Delta\hat{\sigma}_y \hat{L}_z$ which splits linearly polarised polaritons to an extent proportional to their angular momentum. Here $\Delta$  is the strength of the SOI and $\hat{L}_z = -i\hbar \partial_\phi$ corresponds to the angular momentum operator. 

The interaction between particles is governed by the nonlinear operator:
\begin{equation*}
    \hat{\mathcal{H}}_{\rm nl}=\hbar g\begin{pmatrix} |\Psi_{+}|^2 & 0 \\ 0 & |\Psi_{-}|^2 \end{pmatrix} 
\end{equation*}
that accounts for the the interaction between polaritons with parallel spins of the magnitude $\hbar g$. Here we neglect the interaction between polaritons with opposite spins   since it is usually weak \cite{Vladimirova}. Besides, from what follows, we focus on studying the impact of the SOI on the circular persistent current. To that end, we neglect the driven-dissipative effects responsible for the appearance of internal currents in the presence of the local gain-dissipation imbalance \cite{Ostrovskaya2012,Sedov2021}. Which is why we consider the general model of a spinor condensate with the fixed number of particles $N_{\rm pol}$. The  population of the condensate matters in the nonlinear problem adressed in the next section. We assume that $N_{\rm pol}$ can be tuned by the pump power although various approaches to control the polariton-polariton interaction strength via electric field or Feshbach resonance technique are known \cite{Takemura, Savvidis}. 

We are interested in the stationary solutions:
	\begin{equation}\label{eq:2}
		\bm{\Psi}(\phi,t) = \bm{\Psi}(\phi) e^{-i\mu t},
	\end{equation}
where $\mu$ is the eigenfrequency.	
 
The important characteristic of circular supercurrents is the orbital angular momentum (OAM). For a spinor condensate one can define the average OAM for any spin-component individually:
 \begin{equation}\label{eq:3}
 	\ell_{\sigma} =\frac{\langle \hat{L}_z \rangle _{\sigma}} {\hbar N_{\rm pol,\sigma}},
 \end{equation} 
where $N_{\rm pol,\sigma}$ is the total number of the $\sigma$-polarised polaritons ($\sigma$ can stand  for the right or left circular polarisation, $X$- or $Y$-polarisations  etc.). The average angular momentum of the $\sigma$-polarised polaritons can be related to the corresponding density current $j_\sigma(\phi)=i\hbar/\left({2m^\star}\right)\left(\Psi_\sigma\partial_\phi\Psi_\sigma^* - \Psi_\sigma^*\partial_\phi\Psi_\sigma  \right)$ as: 
\begin{equation}\label{eq:4}
    \langle L_z \rangle_\sigma=m^\star \int j_\sigma(\phi)\, d\phi.
\end{equation}
Analogously, the global circulation of the spinor condensate can be characterised by the total OAM per particle $\ell = \hbar^{-1}{\langle \hat{L}_z \rangle} \left/{ N_{\rm pol}}\right.$ which is governed by the density current $j(\phi)=i\hbar/\left({2m^\star}\right)\left((\partial_\phi\bm{\Psi}^\dagger) \bm{\Psi} - \bm{\Psi}^\dagger\partial_\phi\bm{\Psi} \right)$. In what follows the OAM is counted in units of $\hbar$.

The non-interacting condensate confined in an ideal ring supports supercurrents in the form of plane waves. In the absence of SOI, the corresponding wave functions and eigenfrequencies: 
\begin{equation}\label{eq:5}
    \Psi(\phi)=A_k e^{ik\phi}+B_k e^{-ik\phi}, \hspace{20pt} \mu=\frac{\hbar k^2}{2m^\star}
\end{equation}
can be parameterised by the integer $k$ playing the role of the supercurrent vorticity. Here the plane wave amplitudes $A_k$ and $B_k$ are arbitrary numbers which satisfy the normalization condition $\int|\Psi|^2 d\phi= N_{\rm pol}$.  As a result,  the OAM in a uniform ring is not fixed but falls within the range from $-k$ to $k$. 

This flexibility is a consequence of the energy degeneracy of the counter-rotating currents which stems from the time-reversal and the continuous rotational symmetries of the ideal ring. The degeneracy is lifted by the potential defect which breaks rotational symmetry. The amplitudes of the counter-propagating currents are no longer independent but coupled by the delta-barrier. Once the plane wave meets the defect, it scatters particles back feeding the oppositely rotating wave. The stationary solution of the problem is still governed by equation~\eqref{eq:5} although the wave number $k$ is not necessarily integer. 

Tacking into account the discontinuity of the derivative at the defect position one obtains two classes of solutions. The first one is characterised by $k \in \mathbb{Z}$ and has a node at the defect. 
The second one has non-integer $k \notin \mathbb{Z}$ whose values are governed by the transcendental equation:
\begin{equation}\label{eq:6}
  {\hbar k}=\beta m^\star\tan\left(\pi k\right).
\end{equation}
For both solutions $\left|A_k\right|=\left|B_k\right|$. Thus two counter-propagating currents perfectly compensate each other, so that OAM per particle is always zero according to equations~\eqref{eq:3}-\eqref{eq:5}. The defect, whatever small it is, destroys persistent currents in a split-ring geometry.

The situation changes drastically in the spin-orbit coupled spinor condensate. The general solution of the linear problem:
\begin{equation}\label{eq:7}
    \bm{\Psi}(\phi)=\mathbf{h}_{\rm d}\left( C_1 e^{ik_1\phi}+C_2 e^{-ik_2\phi}\right)+ \mathbf{h}_{\rm ad} \left( C_3 e^{ik_2\phi}+  C_4 e^{-ik_1\phi}  \right),
\end{equation}
is governed by the wave numbers $k_{1,2}={m^\star}{\hbar^{-1}}\left(\sqrt{\Delta^2+{2\hbar\mu}/{m^\star}} \mp \Delta\right)$. Here  $\mathbf{h}_{\rm d}=\left(1,-i\right)^\intercal$ and $\mathbf{h}_{\rm ad}=\left(1,i\right)^\intercal$ are the eigenvectors of the spinor part of the master Hamiltonian $\hat{\mathcal{H}}_{\rm l} = \hat{\mathcal{H}}_{\rm 0} + \hat{\mathcal{H}}_{\rm SOI}$. Namely, $\mathbf{h}_{\rm d}(\mathbf{h}_{\rm ad})$  corresponds to the diagonal (anti-diagonal) polarisation. 

In an ideal ring, the periodic boundary conditions require either of $k_{1,2}$ to be integer which yields the energy spectrum $\mu_{1(2)}={\hbar k_{1(2)}^2}\left/{2m^\star}\right. \pm\Delta k_{1(2)}$. As a result, $C_2=C_3=0$ at $k_1\in \mathbb{Z}$ or $C_1=C_4=0$ at $k_2 \in \mathbb{Z}$. 
Herewith, the spinor condensate represents a superposition of two counter-rotating currents with opposite vorticities whose polarisations are orthogonal in the diagonal-antidiagonal basis. Their amplitudes are arbitrary, herewith the total OAM is not fixed but falls within the range from $-k_{1(2)}$ to $k_{1(2)}$. The only exception is the case of a fourfold degeneracy, $\mu_{1}=\mu_{2}$, which is realised at  $\Delta = n\hbar/2m^\star$, where $n \in \mathbb Z$. In this case, $|\ell| < (k_1 + k_2)/2$. 

In contrast to the scalar case, the defect can't inhibit rotation of the spinor condensate. Although the rotation symmetry is violated, the
time-reversal symmetry of the conservative {spin-$1/2$} problem still presumes a twofold degeneracy of each energy level due to the Kramers degeneracy theorem  \cite{Sakurai}. This allows for a flexible choice of the current amplitudes.

The straightforward solution of the problem yields the energy spectrum demonstrated in Fig.~\ref{fig:1}b. The structure of the solution equation~\eqref{eq:7} is defined from the equation:
\begin{equation}\label{eq:8}
    \hbar\sin\left({\pi k_1}\right)\sin\left({\pi k_2}\right) (k_1+k_2) = \beta m^\star \sin\left({\pi(k_1+k_2)}\right).
\end{equation}
The resulting total OAM per particle of the condensate reads:
		\begin{equation}\label{eq:9}
		\ell=\frac{k_1f(k_2)-k_2f(k_1)+\kappa f(k_1)f(k_2)(k_1-k_2)}{f(k_2)+f(k_1)+2\kappa f(k_1)f(k_2)}\times   \frac{|C_{1}|^2-|C_{4}|^2}{|C_{1}|^2+|C_{4}|^2},
		\end{equation}
where $f(k)=\sin^2({\pi k})$ and $\kappa=\frac{2\hbar}{\pi\beta m^\star}$ is governed by the kinetic energy to the defect strength ratio. The value of the net current is governed by the amplitudes of the opposite waves $C_1$ and $C_4$ (or equivalently $C_2$ and $C_3$) which can be excited in arbitrary proportions. 

Therefore, the spinor condensate possesses an intrinsic flexibility in supporting circular persistent currents. Contrary to the scalar case, the defect just affects the current amplitude rather than pulls it up. Besides, it affects the energy spectrum leading to the anti-crossing at the SOI strength $\Delta$ multiple of $\hbar/2m^\star$ which is characterised by the four-fold degeneracy in an ideal ring in the presence of SOI, see Fig.~\ref{fig:1}b. Note that according to equation~\eqref{eq:9} the maximal OAM is independent of the defect strength, in this case.

\subsection*{The nonlinear currents in the spinless split-ring condensate}\label{sec:NonlCurSpinless}

The absence of the circular current in the spinless split-ring condensate discussed in the previous section appeared to contradict the superfluid nature of the polariton condensate. In this section, we demonstrate that the nonlinear interactions allow to maintain persistent currents. We address the nonlinear problem numerically. In particular, we solve the scalar analog of the stationary problem described by equations~\eqref{eq:1},~\eqref{eq:2} iteratively. In what follows we focus on the few lowest energy states whose winding number absolute values are lower than 1.

\begin{figure}[h!]
\centering
    \includegraphics[width=0.75 \linewidth]{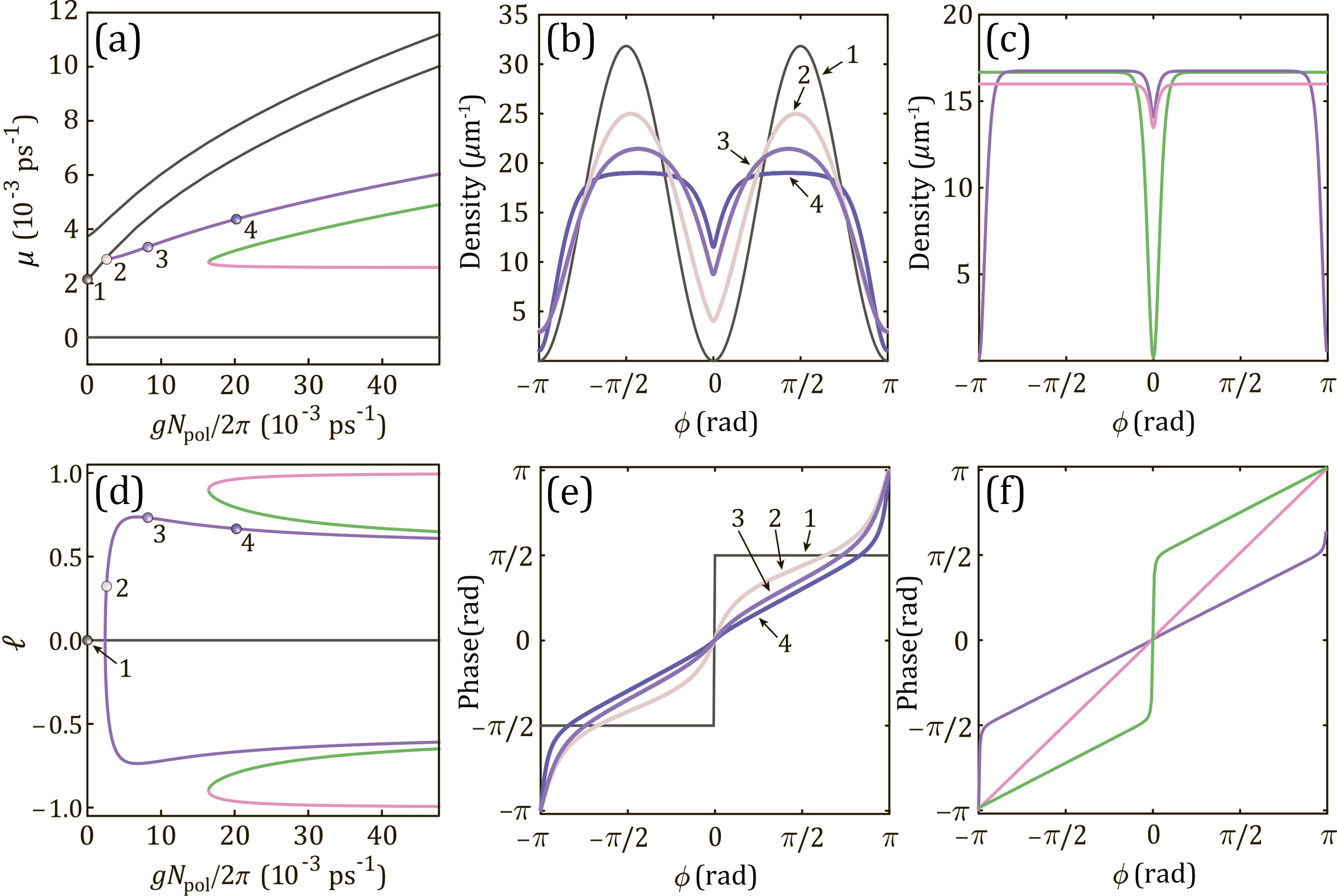}
    \caption{Nonlinear currents states of the spinless split-ring condensate.  (\textbf{a}) and (\textbf{d}) demonstrate appearance of the nonlinear current solutions with the increase of the average  nonlinear blue shift per particle $g N_{\rm pol}/2\pi$. The current states with unit winding number are shown with the color lines while the state with zero OAM $\ell=0$ are grey.  The eigenfrequency counted from the ground state energy is shown in (\textbf{a}) while the average OAM $\ell$ is shown in the panel (\textbf{d}). (\textbf{b}) and (\textbf{e}) show the scenario of the bifurcation of the nonlinear current. (\textbf{c}) and (\textbf{f}) show the azimuthal stucture of the density $|\Psi|^2$ and the phase ${\rm arg}(\Psi)$ of the complex wave function for three lowest nonlinear current states in the large $g$-limit. Namely, at $gN_{\rm pol} = 1.5$~ps$^{-1}$. }
    \label{fig:2}
\end{figure}

A non-interacting split-ring condensate has an even number of density nodes, see equation~\eqref{eq:5}. These states persist in the interacting condensate as well, although the nonlinearity affects their azimuthal density profiles. As in the linear case, the corresponding wave functions have no phase gradient. Herewith, these states sustain no current. The main peculiarity of the nonlinear regime is the presence of the rotating states which supports non-zero circular currents. These states appear at the up-critical strength of nonlinear interactions while the critical value is governed by the defect strength \cite{Seaman2005}.

The existence of the persistent current flowing through the defect can be assigned to the manifestation of a superfluid behaviour \cite{Muller}. In contrast to the non-rotating solutions, the current states typically have either zero or odd number of the density nodes. Besides, they are featured by the presence of the local density dips characteristic of the grey solitons \cite{Carr}.

Figure~\ref{fig:2} shows the bifurcation of the current state from the non-rotating solution with the increase of the nonlinear energy shift. Practically, the interaction strength can be controlled by manipulating the condensate population. In polariton systems this can be achieved by tuning the pump power. Besides, various experimental techniques which allow to control the nonlinearity strength can be implemented \cite{Takemura,Savvidis}. The rotating state  bifurcates from the second energy level at the finite interaction strength. Its average OAM gradually grows starting from zero and approaching $\ell\approx 0.5$ at large $g$. This state is characterised by a grey soliton located at the defect which evolves from the density notch according to the scenario shown in Fig.~\ref{fig:2}b and Fig.~\ref{fig:2}e. At large $g$ [Fig.~\ref{fig:2}c and Fig.~\ref{fig:2}f], the state structure is dominated by the deep density notch resembling a dark soliton on the opposite side from the defect. The corresponding $\pi$-jump of the condensate phase ${\rm arg}(\Psi)$ leads to the half-integer OAM of the state.

In addition, two more current states with equal to one vorticity appear at the strong nonlinearity, see Fig.~\ref{fig:2}a and Fig.~\ref{fig:2}d. They are the half-integer current with the soliton located at the defect position and the vortex state with $\ell\approx 1$ and a weak density dip on the defect. The origin of these states will be explained in the next section.

\subsection*{Classification  of  the  nonlinear  states  of the split-ring spinor condensate} 

The nonlinear self-interaction of the condensate like the SOI in the linear case allows to wind up the condensate even in the presence of a defect. In this section, we study an interplay of these phenomena in a spinor split-ring condensate. As in the scalar case, the nonlinear spinor problem can be addressed with the use of numerical methods. The extensive simulations reveal a plethora of stationary states. They can be classified according to the symmetry of their spinor wave functions.

\subsubsection*{Linearly polarised solutions}\label{item(i)} 

The first group consists of the eigenstates of the spinor part of the Hamiltonian \eqref{eq:1} which obey the same symmetry as the linear solutions. This allows to represent the solution as a product of a spin and a spatial functions: 
\begin{equation}\label{eq:10}
\bm{\Psi}_{\rm d,ad}=\mathbf{h}_{\rm d,ad}\Psi_{\rm d,ad}(\phi),    
\end{equation}
where $\mathbf{h}_{\rm d,ad}$ correspond to the diagonal (antidiagonal) polarisation. Herewith, the solutions of this kind possess a uniform polarisation of the condensate. The spatial wave function $\Psi_{\rm d,ad}(\phi)$ is governed by the following stationary problem which is obtained by substituting equation~\eqref{eq:10} in equation~\eqref{eq:1}:
\begin{equation}\label{eq:11}
	\hbar\mu \Psi_{\rm d,ad} = \left[\hat{{h}}_0 + \hbar g \left|\Psi_{\rm d,ad}\right|^2 \mp  i\hbar\Delta\partial_\phi \right] \Psi_{\rm d,ad} . 
\end{equation}

The states equation~\eqref{eq:10} are connected via the time reversal transformation (up to the global phase) $\hat{\mathcal{T}}=i\hat{\sigma}_y\hat{\mathcal{K}}$, where $\hat{\mathcal{K}}$ is the complex conjugation operator. Therefore, the system preserves a twofold degeneracy of its energy levels.  Note that in contrast to linear case, the degenerate states of the nonlinear problem are unique in the sense that none of the linear combinations $\bm{\Psi} = c_1 \bm{\Psi}_{\rm d}  + c_2 \bm{\Psi}_{\rm ad} $ is a solution.

Note that equation~\eqref{eq:11} is equivalent to the problem of a nonlinear ring-shaped condensate subject to the effective magnetic field  up to the global energy shift proportional to $\Delta^2$. Thus one expects the appearance of a supercurrent which is equivalent to a persistent current of a charge on a ring in the normal magnetic field. The sign at the SOI term in equation~\eqref{eq:11} governs the condensate rotation direction. Therefore, $\ell_{\rm d} = - \ell_{\rm ad}$ while the total current is always finite in the presence of SOI.  Another neat analogy is the problem of a 1D nonlinear condensate in the Dirac comb \cite{Seaman2005}. In the latter case, the presence of the non-zero quasimomentum of the Bloch solution is equivalent to the finite SOI.

\subsubsection*{Spin-flip-parity-symmetric states} \label{item(ii)}
 
The previous section demonstrates that for the spin eigenstates, the current solutions appear due to the spin-orbit coupling. One can ask, whether there exists the state which remains static even in the presence of SOI? The answer is positive. Such a state should obey a symmetry operation upon which $\hat{L}_z$ is odd. This is the case of mirror reflection which inverts the rotation direction. In the 1D case with periodic boundary conditions, the reflection about the defect position is equivalent to the parity transformation $\phi \rightarrow -\phi$. This operation commutes with the SOI term once it is followed by a spin flip $\hat{\mathcal{F}}=\hat{\sigma}_x$. Therefore, it commutes with the linear Hamiltonian, i.e. $\left[\hat{\mathcal{H}_{\rm l}}, \hat{\mathcal{P}}\hat{\mathcal{F}} \right]=0$, where $\hat{\mathcal{P}}$ is the 1D parity inversion operation, $\Psi_\pm(\phi) \rightarrow \Psi_{\pm} (-\phi)$. The nonlinear terms in equation \eqref{eq:1} obey the same symmetry. So, the spinor condensate is characterized by a family of solutions which correspond to the eigenstates of the joined parity and spin-flip operator. 

The $\hat{\mathcal{P}}\hat{\mathcal{F}}$-operator has two eigenstates:
 \begin{equation}\label{eq:12}
 	\bm{\Psi}_{\rm e,o} = \begin{pmatrix} \Psi_0(\phi) \\ \pm \Psi_0(-\phi) \end{pmatrix}.
 \end{equation}
Here $\bm{\Psi}_{\rm e} = \hat{\mathcal{P}} \hat{\mathcal{F}} \bm{\Psi}_{\rm e}  $ is the even and $\bm{\Psi}_{\rm o} = -\hat{\mathcal{P}} \hat{\mathcal{F}} \bm{\Psi}_{\rm o} $ is the odd state. Since $\hat{\mathcal{P}}\hat{\mathcal{F}}$ is a unitary operator whose eigenvalues are $\pm 1$, its eigenstates carry no net-current as long as the corresponding average OAM always vanishes, $\langle \hat{L}_z \rangle = - \langle \hat{L}_z \rangle = 0 $.

The function $\Psi_0(\phi) $ which governs the shape of the $\hat{\mathcal{P}} \hat{\mathcal{F}}$-symmetric states obeys the equations:
\begin{equation}\label{eq:13}
   \hbar \mu \Psi_{0} (\phi) = \left(\hat{{h}}_{0} + \hbar g \left|\Psi_0(\phi)\right|^2 \right)\Psi_{0} (\phi) \pm  \hbar\Delta \partial_\phi \Psi_{0}(-\phi),
\end{equation}
where the plus (minus) sign at the SOI term corresponds to the $\bm{\Psi}_{\rm e}$ ($\bm{\Psi}_{\rm o}$) solution. 

\subsubsection*{The states without specific symmetry}\label{item(iii)}
On the top of the states mentioned above, there are solutions whose spinor wave functions $\bm{\Psi}$ are not subjected to any symmetry. Their common feature is that the opposite circularly polarised condensates occupy the states which originate from different eigenstates of the  scalar nonlinear problem. 

In order to comprehend the structure of such states, consider the SOI-free case $\Delta=0$. The circular polarisations are decoupled in this case, therefore $\Psi_+$ and $\Psi_-$ can occupy different energy levels of the nonlinear scalar problem. Say, $\Psi_+=\Psi_{\rm gr}$ is in the ground state, while $\Psi_-=\Psi_{1/2}$ is in the half-integer current state discussed in  subsection dedicated to the nonlinear currents in the spinless case. Since these states have different energies, the spinor $\bm{\Psi} = \left( \Psi_{\rm gr}, \Psi_{1/2}\right)^\intercal$ is not a stationary solution of the problem equation~\eqref{eq:1}. The spin-orbit interaction, however, couples the spin components which alters their structure. If the coupling is strong enough, it may bring spin components into the resonance (analogously to the scenario typical of synchronisation phenomenon). 
 
\section*{Discussion}
 	
Studying the linearly polarised eigenstates \eqref{eq:10}, we focus on the solutions with the uniform anti-diagonal polarisation $\bm{\Psi}_{\rm ad}$ for which the total OAM is $\ell=\ell_{\rm ad}$. Its time-reversal counterpart is polarised orthogonally and rotates in the opposite direction, $\ell_{\rm d}=-\ell$, which implies  $\Psi_{\rm d} = \Psi^*_{\rm ad}$.

\begin{figure}[h!]
	\centering
	\includegraphics[width=\linewidth]{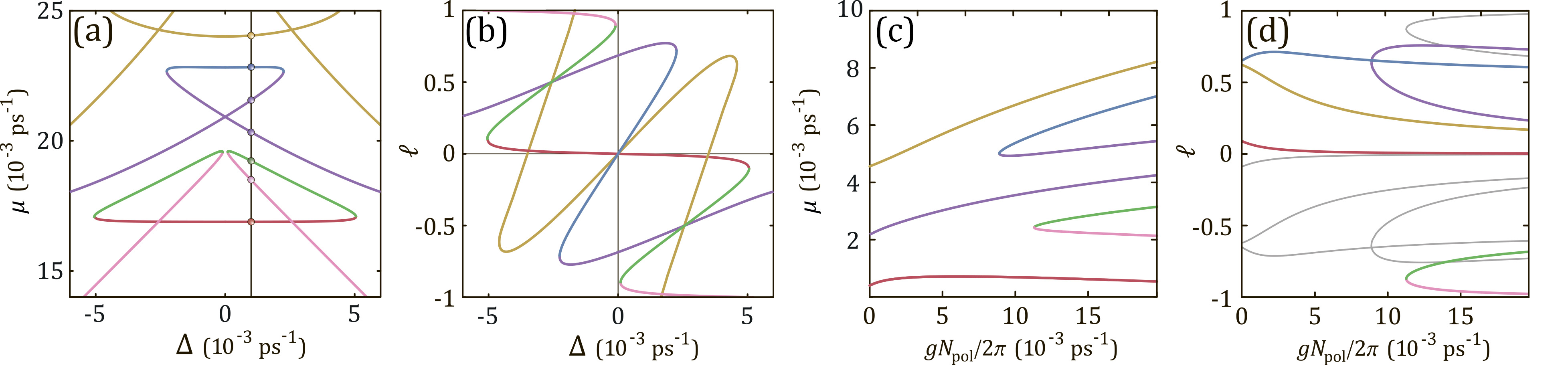}
	\caption{ The nonlinear energy spectrum of the anti-diagonally polarised eigenstates of equation~\eqref{eq:1}. The demonstrated results are obtained by the numerical solution of equation~\eqref{eq:11}. (\textbf{a}) The eigenfrequencies, $\mu(\Delta)$ and (\textbf{b}) the $\ell(\Delta)$-dependence at the fixed average energy shift per particle corresponding to $g N_{\rm pol}=0.1$~ps$^{-1}$. (\textbf{c}) The $\mu(g)$- and the $\ell(g)$-dependencies (\textbf{d}) at the fixed SOI strength $\Delta=10^{-3}$~ps$^{-1}$. The colors are consistent with those shown in (\textbf{a}) and (\textbf{b}) as well as with Fig.~\ref{fig:2} for the purple, green and pink curves. The pale grey curves in (d) show the OAM of the $\bm{\Psi}_{\rm d}$ solutions.  A constant slope of the $\mu(g)$-dependence was subtracted for the convenience purposes.
	The defect strength is $\beta = 0.1$~ps$^{-1}$. 
	The other parameters are the same as in Fig.~\ref{fig:1}b.}
	\label{fig:3}
\end{figure}

Figure~\ref{fig:3} demonstrates the energy spectrum and the corresponding total OAM per particle. The $\Delta$-dependencies calculated at the fixed interaction strength $g N_{\rm pol}=0.1$~ps$^{-1}$ are shown in the panels (a) and (b). The obtained $\ell(\Delta)$-dependence [Fig.~\ref{fig:3}b] reveals an implicit correlation between the presence of SOI and the global rotation of the linearly polarised condensate. In contrast to the spinless problem whose spectrum includes non-rotating states, the linearly polarised spinor condensate rotates with finite OAM at any magnitude of the SOI.

\begin{figure}
	\centering
	\includegraphics[width=1\linewidth]{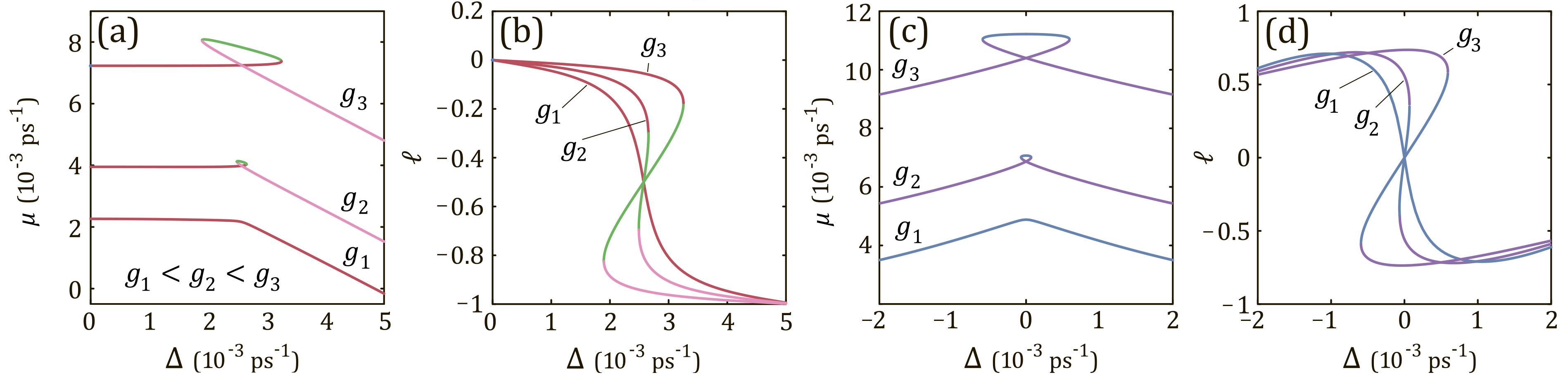}
	\caption{ Formation of the loop structures. The interaction strength increases from $g_1 N_{pol}=10^{-2}$~ps$^{-1}$ to $g_2 N_{pol}=2\times 10^{-2}$~ps$^{-1}$ and finally to $g_3 N_{pol}=4\times 10^{-2}$~ps$^{-1}$. (\textbf{a}) and (\textbf{b}) show the azimuthal variation of the eigenfrequencies $\mu$ and the average OAM per particle $\ell$ of the ground state (red curves). The green curves correspond to the state with a soliton localised at the defect. The pink curves correspond to the vortex-like state with $\ell\approx -1$. (\textbf{c}) and (\textbf{d}) demonstrate the same as in (\textbf{a}) and (\textbf{b}) but for the second currentless energy eigenstate of the spinless problem (blue curves). The line colors match with those used in Fig.~\ref{fig:3}.  The purple curves show the half-integer current states which has a grey soliton on the opposite side from the defect.}\label{fig:4}
\end{figure}

The loop-like structures in Fig.~\ref{fig:3}a demonstrate an interplay between the eigenstates of the scalar problem. In particular, the ground (red curve), half-integer current state (green curve) with the density notch located at the defect and the vortex state (pink curve) are continuously transformed into one another with the variation of the SOI strength. The energy loops are an intrinsic attribute of the nonlinear band structure \cite{Seaman2005,Muller,Chestnov}. Their origin can be understood from Fig.~\ref{fig:4} where their formation with the increase of the $g$-parameter is demonstrated. The analogy with the Bloch problem in the periodic boundary conditions mentioned in the previous section partially explains the loop structure of the spectrum. The similar energy structures were predicted for the atomic BEC loaded in a lattice \cite{Seaman2005}.

The general property of the $\hat{\mathcal{P}}\hat{\mathcal{F}}$-symmetric states is the vanishing of the total OAM. It contrasts to the linearly polarised states which always rotate in the presence of the SOI. The global rotation induced by the SOI is screened by the $\hat{\mathcal{P}}\hat{\mathcal{F}}$-symmetric states because these states support the opposite circular currents in the different spin-components: 
\begin{equation}\label{eq:14}
	\ell_+ = - \ell_-.
\end{equation}
A proof straightforwardly follows from the definitions equations~\eqref{eq:3} and \eqref{eq:12}. The only exception is a special case of $\ell_+ =  \ell_- = 0$. Condition equation~\eqref{eq:14} provides us with an intuition on the origin of the $\hat{\mathcal{P}}\hat{\mathcal{F}}$-symmetric solutions. In the absence of SOI, $\Delta=0$, they are reduced to the states whose circular polarisations occupy the same energy level while the polaritons flow in opposite directions. 

\begin{figure}
	\centering
	\includegraphics[width=1\linewidth]{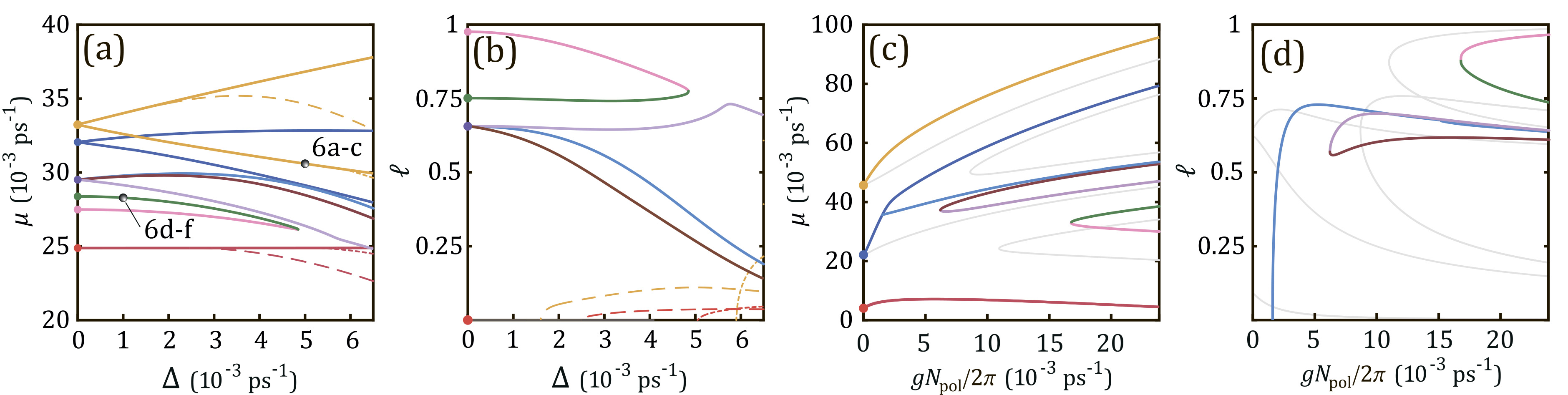}
	\caption{ The energy and the angular momentum of the even  $\hat{\mathcal{P}}\hat{\mathcal{F}}$-symmetric states. (\textbf{a})  The $\mu(\Delta)$-dependence at $gN_{pol}=0.15$~ps$^{-1}$. The dots on the left axis correspond to the eigenenergies of the nonlinear spinless condensate. From top to bottom: the symmetric (yellow) and the antisymmetric (deep blue) with respect to the defect, half-integer current with the soliton located opposite to the defect (purple),  half-integer current with the soliton pinned to the defect (green), vortex-like (pink) and the ground state (red). The solid lines correspond to the state which evolve from the scalar solutions as $\Delta$ increases. The dashed curves show the states which appear at finite $\Delta$ and have no counterparts in the spinless case. 
	(\textbf{b}) the  $\ell(\Delta)$-dependence of the states shown in panel (\textbf{a}).
	(\textbf{c}) The $\mu(g)$-dependence at $\Delta=10^{-3}$~ps$^{-1}$.	The pale grey curves show the energy of the diagonally polarised states from Fig.~\ref{fig:3}.  The constant slope of the $\mu(g)$-dependence has been subtracted.  Only the angular momenta of the $\Psi_+$ polaritons are shown in (\textbf{b}) and (\textbf{d}). The corresponding {{$\ell_-(\Delta)$-dependencies}} are the mirror images of the given curves reflected about the $\ell=0$ axis.}\label{fig:5}
\end{figure}

To be specific, we focus on the even $\hat{\mathcal{P}}\hat{\mathcal{F}}$-symmetric states, $\bm{\Psi}_{\rm e}$. The corresponding energy spectrum is shown in Fig.~\ref{fig:5}. In agreement with qualitative arguments given above, in the absence of SOI, $\Delta=0$, there are six energy states whose winding number absolute values are less than 1. Because of the independence of the spin components, the problem can be reduced to the spinless case discussed above. At finite $\Delta$, the number of solutions increases. They can be divided into three groups.

 \textbf{(i)} The states which propagate no current in both circular polarisations: the corresponding wave functions $\Psi_+$ and $\Psi_-$ demonstrate no phase gradient. The solutions of this kind evolve from the those states whose scalar counterparts are also currentless, namely, from the ground state and both upper energy states shown with the grey curves in Fig.~\ref{fig:2} (see also the red, blue and yellow points on the left horizontal axis of Fig.~\ref{fig:5}a).

\begin{figure*}[h!]
	\centering
	\includegraphics[width=0.75\linewidth]{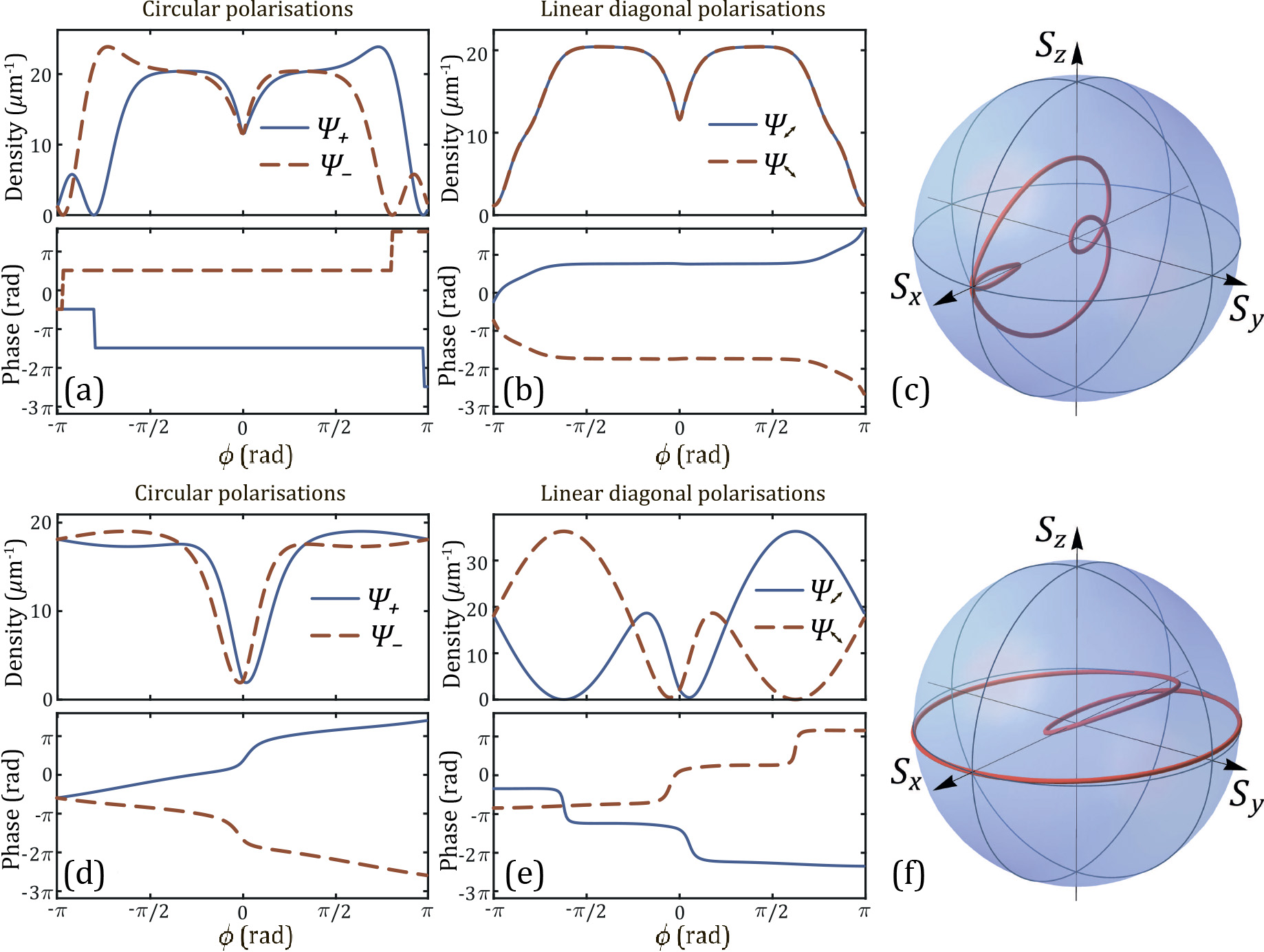}
	\caption{The properties of the $\hat{\mathcal{P}}\hat{\mathcal{F}}$-symmetric states. (\textbf{a})-(\textbf{c})  the state which stems from the currentless symmetric nonlinear solution. Its position on the energy diagram is indicated in Fig.~\ref{fig:5}a. (\textbf{d})-(\textbf{f})  the state which stems from the half-integer current state with a grey soliton at the defect position.}\label{fig:6}
\end{figure*}

An example of such a state is demonstrated in Fig.~\ref{fig:6}a-c. The density distributions of the $\Psi_+$- and $\Psi_+$- components are connected via the 1D parity inversion, see panel (a). Note that the wave functions of the diagonally polarised polaritons are complex conjugate, $\Psi_{\rm d} = \Psi_{\rm ad}^*$ because of the time-reversal symmetry. As a results, the currents of diagonally polarised polaritons flow in opposite directions, $\ell_{\rm d} = - \ell_{\rm ad}$, see Fig.~\ref{fig:6}b. In contrast to the states considered in the previous subsection, these solutions have nontrivial polarisation properties. Their polarisation can be conveniently characterised using the Stokes vector representation:
\begin{equation}\label{eq:15}
	\mathbf{S}=\frac{1}{2} \bm{\Psi}^\dag \cdot \bm{\sigma} \cdot \bm{\Psi}.
\end{equation}
The spatial dependence of the Stokes vector along the ring circumference can be visualised as a closed curve within the Poincare sphere, see Fig.~\ref{fig:6}c. The Stokes vector belongs to the plane $\left(S_x,S_z\right)$ which follows straightforwardly from the condition $\left|\Psi_{\rm d} \right| = \left|\Psi_{\rm ad} \right|$.

\textbf{(ii)} The second group is represented by the states with  opposite currents in the orthogonal circular polarisations. They stem from those states which have the current in the scalar case (see the purple, green and pink points on the left horizontal axis of Fig.~\ref{fig:5}a and the caption therein). 

The phase and the density profiles of such states are illustrated in Fig.~\ref{fig:6}d and Fig.~\ref{fig:6}e. 
The wave functions of the diagonally polarised polaritons are characterised by two solitons each. The resulting spatial-dependence of the Stokes vector is shown in Fig.~\ref{fig:6}f. There is a common symmetry in the behaviour of the $\mathbf{S}$-vector. In such a representation, the trajectory passed by the  Stokes vector is simmetric to the mirror reflection with respect to any Cartesian axis.

\textbf{(iii)} The third group of solutions is represented by the states which bifurcate from the solutions of the first group as $\Delta$ increases. They are shown by the dashed lines in Fig.~\ref{fig:5}a and Fig.~\ref{fig:5}c. The circular polarisation components of these states always carry opposite currents in contrast to the states from which they bifurcate.

\section*{Conclusions}

We have addressed the model of a spinor polariton condensate loaded in a one-dimensional ring split by a delta-function potential. We find an analytical solution of the spinor  Schr\"{o}dinger equation in the presence of SOI and show that a defect, whatever large it is, can not stop the condensate rotation in general. In the nonlinear regime, the states with the intrinsic circular current appear. They are formed even in the absence of SOI. Their structure is characterized by the presence of the soliton-like density deeps accompanied by the local jumps of the phase which are responsible for the fractional OAM per particle of the condensate. The coupling between the pseudospin and the global rotation of the condensate affects its energy spectrum and it alters the condensate rotation. We distinguish between two main types of stationary states of the spinor condensate. The SOI acts as a torque on the uniformly linearly polarised condensate which inherits an additional rotation from the interaction with its pseudospin. However, the states which obey the symmetry of simultaneous spin-flip and the mirror reflection can resist against the SOI-induced rotation. This regime is achieved by supporting the oppositely directed currents of spin-up and spin-down condensates. This work paves the way to engineering of the energy spectra of polariton qubits and shows how the quantum state of a qubit based on the split-ring polariton condensate can be controlled by the polarisation of the condensate.

\section*{Acknowledgements}

We acknowledge support from Rosatom within the Roadmap on quantum computing.
The work is supported by the Westlake University, Project 041020100118 and Program 2018R01002 funded by the Leading Innovative and Entrepreneur Team Introduction Program of Zhejiang Province. The support from RFBR Grant 21-52-10005, from the Grant of the President of the Russian
Federation for state support of young Russian scientists No. MK-5318.2021.1.2 is also acknowledged. A.K. acknowledges the Saint-Petersburg State University for the financial support (research grant No 91182694). I.C. acknowledges fruitful discussions with A. Nalitov.

\section*{Author contributions}

A.K. proposed the idea. S.D., I.C. have developed the idea and the formalism. S.D., I.C. and K.K. performed all calculations. All authors contributed to discussions. S.D., I.C. and A.K. have written the paper. All authors reviewed the manuscript.

\section*{Competing interests}
The authors declare no competing interests.

\end{document}